\PassOptionsToPackage{pdfpagelabels=false}{hyperref}
\documentclass[fleqn,usenatbib]{mnras}

\usepackage{mathptmx}

\usepackage[T1]{fontenc}
\usepackage{aecompl}

\usepackage{graphicx}	
\usepackage{amsmath}	
\usepackage{amssymb}	

\usepackage{color}
\usepackage{ulem}

\title[Disk origin of emission lines in TDE PTF09djl]{Disk origin of broad optical emission lines 
of the TDE candidate PTF09djl}

\author[Liu et al.]{F.K. Liu,$^{1,2}$\thanks{E-mail: fkliu@pku.edu.cn (FKL)}
Z.Q. Zhou,$^{1}$ 
R.Cao,$^{1}$  
L.C. Ho,$^{2,1}$ 
S. Komossa$^{3}$
\\
$^{1}$Department of Astronomy, School of Physics, Peking University, Beijing 100871, China\\
$^{2}$Kavli Institute for Astronomy and Astrophysics, Peking University, Beijing 100871, China\\
$^{3}$Max-Planck-Institut f\"ur Radioastronomie, Auf dem H\"ugel 69, 53121 Bonn, Germany
}

\date{Accepted XXX. Received YYY; in original form ZZZ}

\pubyear{2017}

\begin{document}
\label{firstpage}
\pagerange{\pageref{firstpage}--\pageref{lastpage}}
\maketitle

\begin{abstract}
An otherwise dormant supermassive black hole (SMBH) in a galactic nucleus flares up when it tidally disrupts
a star passing by. Most of the tidal disruption events (TDEs) and candidates discovered in the optical/UV
have broad optical emission lines with complex and diverse profiles of puzzling origin.
In this Letter, we show that the double-peaked broad H$\alpha$ line of the TDE candidate PTF09djl 
can be well modelled with a relativistic elliptical accretion disk and the peculiar substructures  
with one peak at the line rest wavelength and the other redshifted to about $3.5\times 10^4 \, {\rm km
\; s^{-1}}$ are mainly due to the orbital motion of the emitting matter within the disk plane of large
inclination $88\degr$ and pericenter orientation nearly vertical to the observer. The accretion disk has 
an extreme eccentricity $0.966$ and semimajor axis of 340 BH Schwarzschild radii. The 
viewing angle effects of large disk inclination lead to significant attenuation of He emission lines originally 
produced at large electron scattering optical depth and to the absence/weakness of He emission lines in 
the  spectra of PTF09djl. Our results suggest that the diversities of  line intensity ratios among the line 
species in optical TDEs are probably due to the differences of disk inclinations.
\end{abstract}




\section{Introduction} \label{sec:intro}

A star is tidally disrupted when it wanders closely by a supermassive black hole (SMBH) in galactic nucleus 
\citep{hil75,ree88}.  About 30-60 stellar tidal disruption events (TDEs) and TDE candidates have been 
observed in the X-rays, UV and optical \citep[for recent reviews]{kom15,auc17}. Few TDEs 
discovered in X-ray show optical emission lines in spectra, but those discovered in the optical have 
strong broad optical emission lines \citep{kom08,gez12,wan12,hol14}. The broad emission lines of the 
optical/UV TDEs and candidates  are complex, asymmetric and of puzzling origin 
\citep{gez12,gas14,gui14,str15,koc16,rot16} and the peculiar spectral characteristics raise skepticism 
on the identification of the optical/UV transients as TDEs \citep{sax17}. 

The optical spectra of the TDE candidate PTF09djl have strong and double-peaked H$\alpha$ emission line
with one peak at the rest wavelength of the line and the other redshifted by about $3.5\times 10^4\, {\rm 
km\;s^{-1}}$ \citep[][see also Fig.~\ref{fig:exa}]{arc14}. The line structure is reminiscent of the 
double-peaked line profiles in active galactic nuclei (AGNs) which are usually explained with disk models 
\citep{che89,era95}. When a circular disk model was applied to the double-peaked
H$\alpha$ profiles, a bulk motion of velocity about $1.5 \times 10^4 \, {\rm km \; s^{-1}}$ has to be 
included to shift the model profiles to the red from their original position to fit it to the observed spectra 
\citep{arc14}. A post-merged BH can obtain a recoiling velocity up to $5000\; {\rm km \, s^{-1}}$ at 
coalescence of two black holes because of gravitational radiations \citep{lou11}. A SMBH 
may have a bulk velocity $1.5 \times 10^4 \, \; {\rm km\, s^{-1}}$, if it is a component of  a BH binary 
of separation $\leq 200 r_{\rm S}$ with $r_{\rm S}$ the BH Schwarzschild radius \citep{liu14}, smaller than
the required disk size \citep{arc14}. A BH binary of mass $10^{6.3} M_\odot$ (with $M_\odot$ the solar 
mass) and separation $\leq 200 r_{\rm S}$ has an orbital period less than $6\; {\rm d}$. Dramatic 
variations of the bulk velocity would be expected during the 60-d spectral observational campaign 
of PTF09djl, inconsistent with the observations. It was suggested that the geometry of the broad-line region 
may be more complex than a circular disk \citep{arc14}. 

Hydrodynamic simulations of stellar tidal disruptions show that the circularization of bound stellar debris
is due to the self-interaction of debris streams because of the relativistic apsidal precession, and that the 
accretion disk would have a large eccentricity except for cases with orbital pericenter of star about the 
BH radius \citep{eva89,shi15,bon16,hay16}. Broad emission lines originating in highly eccentric 
disk would have  distinctive profiles.

In this Letter, we suggest that the broad emission lines of the spectra of PTF09djl originate in an eccentric 
accretion disk as expected from hydrodynamic models for TDEs. We show that the peculiar broad 
double-peaked profiles of the H$\alpha$ emission line can be well reproduced with a relativistic elliptical 
disk model of semimajor axis $340 r_{\rm S}$ and eccentricity $0.966$. The disk plane 
inclines by an angle $88\degr$ and has pericenter orientation nearly vertical to the observer.

\section{The spectral data of PTF09djl}\label{sec:data}

\begin{figure*}
\begin{center}
\includegraphics[width=2\columnwidth]{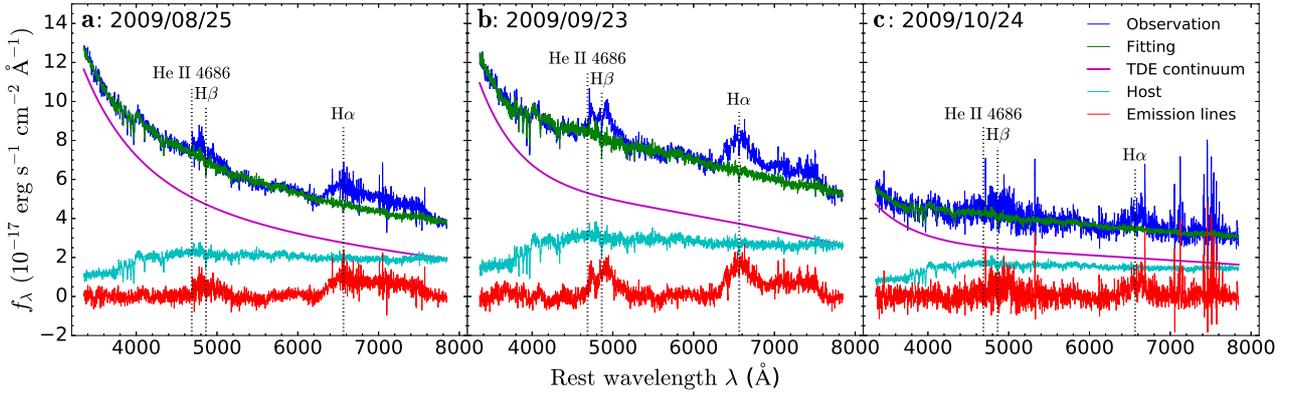}
\caption{Spectral decomposition and emission-line spectra of PTF09djl; the UT dates  of the observations 
are given at the top of each panel. The spectrum was fit with a featureless TDE continuum, globally modelled 
with a third-order polynomial (magenta),  and host galaxy starlight (light blue).  The pure emission-line residual 
spectrum is shown in red.  Balmer emission lines and the expected location of He~II~$\lambda$4686 are 
identified with vertical dashed lines.
\label{fig:exa}}
\end{center}
\end{figure*}

The TDE candidate PTF09djl was discovered in a redshift $z=0.184$ E+A galaxy in the Palomar Transient Factory  
(PTF) survey on 2009 July 24,  and follow-up optical spectra were obtained with the Low Resolution Imaging 
Spectrometer (LRIS) mounted on the Keck~I 10~m telescope for the transient on 2009 August 25, September 
23 and October 24, and for the host galaxy on 2013 May 9\footnote{The data are available online through the 
Weizmann Interactive Supernova data REPository \citep[WISeREP;][]{yar12} at 
http://www.weizmann.ac.il/astrophysics/wiserep} \citep{arc14}. We corrected the spectra for a Galactic 
extinction of $A_{\rm V} = 0.049\, {\rm mag}$ and $R_{\rm V} = 3.1$ \citep{car89,sch11}. The three spectra of 
PTF09djl  and the spectrum of the host galaxy are shown in Fig.~\ref{fig:exa}. The Balmer lines are 
extremely broad and prominent and the transient event can be regarded as the H-dominated in the 
sequence of He- to H-rich TDEs \citep{arc14}. 

We globally model the continuum of the TDE using a third-order polynomial by fitting to several line-free regions 
of the TDE spectrum, plus a scaled spectrum of the host galaxy. The scale factor of the host spectrum is 
determined by matching regions with strong stellar features and without emission lines. Fig.~\ref{fig:exa} 
gives our spectral decomposition procedure for the three spectra and the residual emission-line spectra of 
PTF09djl after Galactic extinction correction and subtraction of the continuum and host galaxy starlight.
H$\alpha$ is strong and double-peaked with the blue peak at the rest wavelength of the line and the red 
one redshifted by about $3.5\times 10^4 \, {\rm km \; s^{-1}}$ at all epochs. H$\beta$ line is also prominent. 
He emission lines are weak or absent from the spectra.

\section{Modeling the broad emission lines}\label{sec:model}

\subsection{Accretion disk in TDEs}\label{sec:disk}

The bound debris after disruption of star is circularized due to the interaction of the outflowing 
and inflowing streams because of the relativistic apsidal precession and the location of the interaction is 
\begin{equation}
    r_{\rm cr} \simeq {(1+e_{\rm mb}) r_{\rm p} \over 1 - e_{\rm mb} \cos{(\Omega/2)}} \simeq {2 x_{\rm 
    p} r_{\rm S}  \over \delta + 2 \sin^2(\Omega/4)} 
    \label{eq:rcros}
\end{equation}
\citep{dai15}, where $r_{\rm p} = x_{\rm p} r_{\rm S}$ with $x_{\rm p} \la 23.54 r_* m_*^{-1/3} M_6^{-2/3}$
is the orbital pericenter of the bound stream, $e_{\rm mb} = 1 - \delta$ with $\delta \simeq 8.49\times 
10^{-4} r_*^{-1} m_*^{2/3} M_6^{1/3}  x_{\rm p}$ is 
the eccentricity of the most-bound stellar debris (with $M_{\rm BH} = 10^{6}M_\odot  M_{\rm 6}$  the  
mass of the BH;  $R_*= r_* R_\odot$ and $M_*=m_* M_\odot$, respectively, the star's radius and mass;
$R_\odot$ the radius of the Sun) and $\Omega$ 
is the instantaneous de Sitter precession at periapse of the most-bound stellar debris after tidal disruption, 
$\Omega = 3\pi r_{\rm S} /[r_{\rm p} (1+ e_{\rm mb})] \simeq 3\pi  /(2  x_{\rm p})$.

The size and orientation of the elliptical accretion disk are determined by the location of the self-interaction 
of streams. Provided that the collision is completely inelastic and the outgoing and incoming streams have 
similar mass, the circularized stellar debris forms an accretion disk of semimajor axis 
\begin{equation}
    a_{\rm d} \simeq {r_{\rm cr} \over 2 \sin^2(\theta_{\rm c}/2)} {1\over   1 + (r_{\rm cr}/ 2 a_{\rm mb}) 
    \cot^2(\theta_{\rm c}/2)} 
    \label{eq:dsize}
\end{equation}
\citep{dai15}, where $a_{\rm mb} \simeq r_{\rm t}^2 /2R_* $ with $r_{\rm t} =  R_* \left(M_{\rm BH} 
/ M_*\right)^{1/3}$ the stellar tidal  disruption radius \citep{gui13,hay13} is the orbital semimajor axis of 
the most bound stellar debris, and $\theta_{\rm c}$ is the stream-stream intersection angle
\begin{equation}
    \cos(\theta_{\rm c}) = {1- 2 \cos(\Omega/2)e_{\rm mb} + \cos(\Omega)e_{\rm mb}^2 \over 1 - 2 \cos(\Omega/2)
    e_{\rm mb} + e_{\rm mb}^2} ,
\end{equation}
or
\begin{equation}
    \sin(\theta_{\rm c}/2) \simeq  {2  \sin(\Omega/4) \over  \sqrt{\delta^2 +4\sin^2(\Omega/4)}} 
    \cos(\Omega/4) \simeq \cos(\Omega/4) .
    \label{eq:intsec}
\end{equation}
Equation~(\ref{eq:dsize}) together with Equations~(\ref{eq:rcros}) and (\ref{eq:intsec}) gives
\begin{equation}
a_{\rm d} \simeq  {2  x_{\rm p}   \over 2\delta  +  \sin^2(\Omega/2)} r_{\rm S}
   \label{eq:thdisk}
\end{equation}
with $\Omega \simeq {3 \over 2} \pi  x_{\rm p}^{-1}$ and $\delta \simeq 8.49\times 10^{-4} 
m_*^{-2/15} M_6^{1/3}  x_{\rm p}$ with $x_{\rm p} \la 23.54 m_*^{7/15} M_6^{-2/3}$, where we 
used $r_* \simeq  m_*^{1-\zeta}$ with $\zeta\simeq 0.2$ for $0.1  \leq m_* \leq 1$ \citep{kip94}.

Neglecting the transfer and redistribution of angular momentum among the shocked plasma, we have 
an eccentricity of the debris disk 
\begin{equation}
e_{\rm d} \simeq \left(1-  {(1+e_{\rm em})r_{\rm p}\over a_{\rm d}}\right)^{1/2} \simeq   \left[\cos^2
\left({\Omega \over 2}\right) - 2 \delta\right]^{1/2}  .
\label{eq:decc}
\end{equation}


\subsection{A disk model for broad emission lines in TDEs}
  
We assume that the broad emission lines in TDEs originate in the elliptical disk given in Sec.~\ref{sec:disk}. 
No observation in hard X-rays was made for PTF09djl, but the survey of TDE candidates  in the archive of {\it 
Swift Burst Alert Telescope} ({\it BAT}) shows that hard X-ray emissions should be ubiquitous in un-beamed TDE
candidates \citep{hry16}. Because the coronal hard X-ray source originates due to the magnetic 
reconnection in the poloidal field lines anchored to the ionized accretion disk, the coronal materials 
should move nearly radially along with the disk material and become radially extended. The configuration 
of coronal X-ray source in TDEs is different from that in AGNs, which is compact.

The calculations of radiative transfer indicate that an ionized optically thick accretion disk irradiated 
by hard X-ray source will produce strong optical emission lines when the ionization parameter is low to 
intermediate \citep{gar13}. The reflection line emissivity of accretion disk at frequency $\nu_{\rm 
e}$ in the frame of the emitter irradiated by X-ray source of finite radial extent $r_{\rm br}$ is
a broken power law in radius $r$ except for the regions near BH horizon 
%
\begin{equation}
I_{\nu_{\rm e}} = \left\{ 
   \begin{array}{r@{\quad}l}
   {\epsilon_0  c \over 2 (2\pi)^{3/2} \sigma} \left({\xi\over \xi_{\rm br}}\right)^{-\alpha_1}
   \exp{\left[-{(\nu_{\rm e}-\nu_0)^2 c^2 \over 2 \nu_0^2 \sigma^2}\right]} & {\rm for} \, 
   \xi \leq \xi_{\rm br} \\
 \\
   {\epsilon_0  c \over 2 (2\pi)^{3/2} \sigma} \left({\xi\over \xi_{\rm br}}\right)^{-\alpha_2}
   \exp{\left[-{(\nu_{\rm e}-\nu_0)^2 c^2 \over 2 \nu_0^2 \sigma^2}\right]} & {\rm for} \,
   \xi > \xi_{\rm br} 
\end{array} \right. 
\end{equation}
with $\alpha_1 \sim 0$ for  corona of constant radial distribution and 
$\alpha _2 \simeq 3$ for $\xi > \xi_{\rm br}$ \citep{wil12,gon17}, where $\xi= r/r_{\rm S}$ and $\xi_{\rm 
br}=r_{\rm br}/r_{\rm S}$, $\epsilon_0$ is a constant,  $\nu_0$ is the line rest frequency and $\sigma$ 
is a velocity dispersion of the local line broadening due to thermal and turbulent motions. A Gaussian 
local line broadening is assumed \citep{era95}. 

\begin{figure}
\begin{center}
\includegraphics[width=\columnwidth]{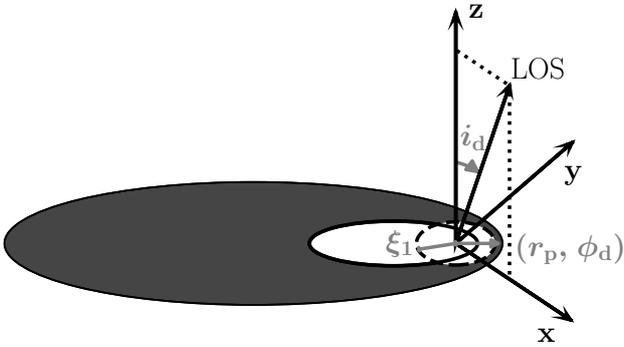}
\caption{The geometry and coordinate system used in 
the profile calculations. The $z$-axis is along the rotation axis of the accretion disk, and the accretion disk 
lies within the $xy$-plane.  The LOS of the observer at infinity is in the $xz$-plane and makes an 
inclination angle $i_{\rm d}$ to the $z$-axis. 
\label{fig:cor}
}
\end{center}
\end{figure}

The observed specific flux at frequency $\nu$ is given by
\begin{equation}
f_\nu  =  {r_{\rm S}^2  \cos{i_{\rm d}} \over d^2} \int  \int            
            I_{\nu_{\rm e}} D^3(\xi,\phi)\; \psi(\xi,\phi) \xi \, 
            \mathrm{d}\xi \, \mathrm{d}\phi\, 
\label{metheq:line}
\end{equation}
\citep{era95,def16}, where $d$ is the luminosity distance to the source, $i_{\rm 
d}$ is the disk 
inclination angle with respect to the line-of-sight (LOS), $\phi$ is the azimuthal angle around the disk 
with respect to the projected LOS in the disk plane and the function $\psi(\xi,\phi)$ describes the 
effects of curved trajectories of light rays \citep{bel02,def16}. Fig.~\ref{fig:cor} shows the geometry and 
coordinate systems. 

%
%

In Equation~(\ref{metheq:line}), $D(\xi,\phi) \equiv \nu/\nu_{\rm e}$ is the Doppler factor and 
describes the effects of gravity and the motion of the emitting particles on the energies of the 
emitted photons \citep{era95,bel02,def16}. The motion of particles in the disk plane is described 
with the generalized Newtonian potential in the low-energy limit \citep{tej13}, which gives the radial 
and azimuthal velocities of the emitting particles in the source frame
\begin{eqnarray}
{1\over c}{\mathrm{d}{r}\over \mathrm{d}{t}} &=&   \left( 1 - {1 \over \xi}\right) \sqrt{2 {E_{\rm G} \over 
       c^2} + {1 \over \xi} - {h_{\rm G}^2 \over r_{\rm S}^2 c^2 }{1\over \xi^2} \left(1 - {1 \over \xi}\right)} ,\\
 {r\over c} { \mathrm{d} {\phi}\over \mathrm{d}{t}}&=&  {h_{\rm G}\over r_{\rm S} c} {\xi - 1 \over \xi^2} 
\end{eqnarray}
with $h_{\rm G}$  and $E_{\rm G}$, respectively, the orbital specific angular momentum and energy. The
generalized Newtonian potential can reproduce exactly the radial dependences of 
the orbital specific binding energy and angular momentum \citep{tej13}.

The eccentric disk is assumed to consist of nested elliptical annuli of constant eccentricity in  radius. 
The particle trajectories in a given elliptical annulus of semimajor axis $a$ and eccentricity $e$ are given by 
$r = a (1 - e^2)  [1+ e \cos(\phi - \phi_{\rm d})]^{-1}$,
where $\phi_{\rm d}$ is the disk orientation and $\phi_{\rm d}  = 0^\circ$ when the pericenter points to the 
observer. When $a$ and $e$ are specified, $h_{\rm G}$ and $E_{\rm G}$ are fixed
\begin{eqnarray}
{h_{\rm G} \over r_{\rm S} c} & = & {(1-e^2) x_{\rm a}  \over \sqrt{2 (1-e^2) x_{\rm a}   - 3 - e^2}} , 
     \label{eq:ang}\\
{E_{\rm G} \over c^2} & = & -{1 \over 2} {x_{\rm a}(1-e^2) - 2 \over x_{\rm a} \left[2 x_{\rm a}(1-e^2) - 
	3-e^2\right]} ,
\label{eq:eng}
\end{eqnarray}
where $x_{\rm a}=a/r_{\rm S}$.


\subsection{Fitting the double-peaked H$\alpha$ profiles of PTF09djl} \label{spec:fit}

Provided pericenter $x_{\rm p}$, the size ($a_{\rm d}/r_{\rm S}$) and eccentricity ($e_{\rm d}$) of 
debris accretion disk are computed with Equations~(\ref{eq:thdisk}) and (\ref{eq:decc}), respectively. 
An inner disk edge $2r_{\rm S}$ is adopted because the accretion disk has an extreme eccentricity
($e_{\rm d} \simeq 0.966$; see Section~\ref{sec:res}) and the orbits of disk fluid elements are 
nearly parabolic, for which the marginal stable orbit for Schwarzschild black hole is about the marginal 
bound orbit $2r_{\rm S}$ and fluid elements passing through $2r_{\rm S}$ fall freely on to the BH. 
The line-emitting region lies between radii $\xi_1$ and $\xi_2 = (1+e_{\rm d}) a_{\rm d}/ r_{\rm S}$. 
We calculate the model line profiles for a 
large parameter space (i.e. $0 \leq i_{\rm d} \leq \pi$, $0\leq \phi_{\rm d} < 2\pi$, $1 \leq x_{\rm p} 
\leq 50 $, $2\leq \xi_1 \leq 70$, $\xi_1 \leq \xi_{\rm br} \leq \xi_2$ and $10^3  \, {\rm km\; s^{-1}} 
\leq \sigma \leq 2\times 10^4 \, {\rm km \; s^{-1}}$) and jointly fit the observed profiles of the 
first two spectra with shared periapse ($x_{\rm p}$) and velocity dispersion ($\sigma$)
with the least-squares method ($\chi^2$). The break radius $\xi_{\rm br}$ might be less than the 
inner radius $\xi_1$ of the emission line region but cannot be constrained observationally. Therefore, 
the calculations are limited to $\xi_{\rm br} \geq \xi_1$. Because the third spectrum has a
low signal-to-noise ratio, the line profile is fitted with the averages $x_{\rm p}$, $\sigma$, $\phi_{\rm 
d}$ and $i_{\rm d}$ obtained with the first two spectra. 

They are calculated with the dimensionless radii ($\xi$, $\xi_1$ and $\xi_{\rm br}$),
orbital pericenter $x_{\rm p}$,  and disk semimajor axis $x_{\rm a}$, but Equations~(\ref{eq:thdisk}) and 
(\ref{eq:decc}) show that the model line profiles depend  explicitly and weakly on the BH mass 
(but nearly independent of the mass of star). The BH mass is $\log(M_{\rm BH}/M_\odot) = 
6.55^{+0.58}_{-0.77}$ estimated with the $M_{\rm BH}$-$M_{\rm bulge}$ relation \citep{arc14} 
and $\log(M_{\rm BH}/M_\odot) = 5.82^{+0.56}_{-0.58}$ with the $M_{\rm BH}$-$\sigma$ relation 
between the bulge of the host galaxy and the mass of the SMBH \citep{wev17}. Both masses 
are consistent with each other within the uncertainties and we adopted the average $M_{\rm 
BH} \simeq 2.1\times 10^6 M_\odot$ in this Letter. A similar BH mass with quite large uncertainties is 
obtained by fitting the H$\alpha$ line profiles. 

\begin{figure}
\includegraphics[width=\columnwidth]{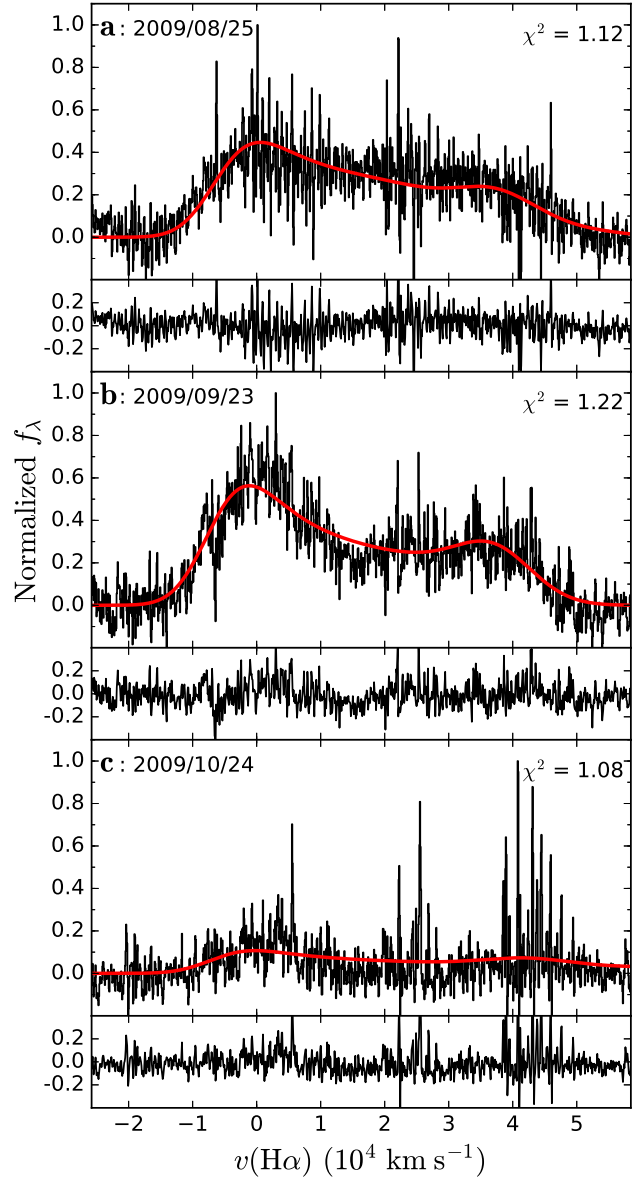}
\caption{Normalized velocity profiles of H$\alpha$ and fits of the model spectra for different epochs (panels
a-c). The UT date is given at the top left of each panel.  The observed profiles (black) are double-peaked and 
are well reproduced with the elliptical disk model (red). The residuals of the spectra after subtraction of the fit 
are given at the bottom of each panel. 
\label{fig:halpha}
}
\end{figure}

\section{Results}
\label{sec:res}

The best-fitting models of the three H$\alpha$ spectra are shown in Fig.~\ref{fig:halpha}, and the 
best-fitting values of the model parameters and their associated errors are given in Table~\ref{tab:diskpar}.  
Table~\ref{tab:diskpar} also gives the reduced $\chi^2$ of the best fits, which is calculated with 
respect to the averaged noises over the closest regions of the emission lines after subtraction of the 
spectral feature. The uncertainties of the fitting parameters at 90\% confidence level are obtained
with the Markov chain Monte Carlo (MCMC) methods.

\begin{table*}
\small
\centering
\caption{Best-fitting disk parameters for the double-peaked H$\alpha$ emission line. \label{tab:diskpar}}
\begin{tabular}{ccccccccccc}
\hline
Date &$r_{\rm p}$($r_{\rm S}$) & $\phi_{\rm d}$($^\circ $) &$i_{\rm d}$($^\circ$) &$a_{\rm d}$($r_{\rm S}$) 
&$\xi_1$ & $r_{\rm br}$ ($r_{\rm S}$) &$e_{\rm d}$ & $\sigma$($\rm km \; s^{-1}$)&reduced-$\chi^2$\\
\hline 
\hline
2009/08/25 & $11.38^{+0.36}_{-1.54}$&$73.59^{+1.34}_{-1.44}$&$87.45^{+1.17}_{-7.29}$
&339.7$^{+20.3}_{-86.3}$&29.52$^{+0.30}_{-5.28}$&$29.53^{+2.41}_{-5.28}$
&0.9659$^{+0.0010}_{-0.0041}$&4713$^{+199}_{-351}$&1.12\\
2009/09/23 & $11.38^{+0.36}_{-1.54}$&70.88$^{+1.55}_{-0.29}$&$88.90^{+0.97}_{-0.30}$&339.7$^{+20.3}_{-86.3}$
&$32.87^{+0.94}_{-5.47}$ &$32.89^{+13.49}_{-5.47}$&0.9659$^{+0.0010}_{-0.0041}$&4713$^{+199}_{-351}$&1.22\\
2009/10/24 & $11.38^{\rm a}$&72.23$^{\rm a}$&88.17$^{\rm a}$&$339.7^{\rm a}$&$20.60^{+5.54}_{-4.98}$
&$41.35^{+12.67}_{-14.92}$&0.9659$^{\rm a}$&4713$^{\rm a}$&1.08\\
\hline
\end{tabular}
\begin{flushleft}
{\it Note}. -- $^a$fixed to the average of the values obtained with the first two spectra.
\end{flushleft}
\end{table*}

The modelling of the three H$\alpha$ spectra of PTF09djl shows that the double-peaked H$\alpha$ line 
profiles with one peak at the line rest wavelength and the other extending to redshift about $3.5\times 
10^4 \, {\rm km\; s^{-1}}$ can be well fitted with a relativistic elliptical disk model without bulk motion. 
Table~\ref{tab:diskpar} shows that the accretion disk has a semimajor $a_{\rm d} \simeq 339.7
r_{\rm S}$ and eccentricity $e_{\rm d} \simeq 0.966$, following circularization of streams with 
orbital pericenter $r_{\rm p} \simeq 11.38 r_{\rm S}$.  During the spectral observations, the  radial 
extent of the hard X-ray source or corona slightly increases from $r_{\rm br} \simeq 29.5 r_{\rm S}$ to 
$32.9 r_{\rm S}$. The contributions of the inner disk regions covered with the extended corona to 
the H$\alpha$ line flux is negligible as the effective size of the regions is
$\Delta{r} = (\xi_{\rm br} - \xi_1) r_{\rm S} \sim 0 $. 
The accretion disk is highly inclined with nearly constant inclination angle $i_{\rm d} \simeq 
88\degr$. The pericenter of the accretion disk is orientated with nearly constant
angle $\phi_{\rm d} \simeq 72\degr$ relative to the observer.
An elliptical disk of high inclination and pericenter orientation nearly vertical to the
observer leads to the formation of the broad (with full width at half-maximum, ${\rm FWHM} \sim 
4\times 10^4 \, {\rm km\; s^{-1}}$) and asymmetric double-peaked H$\alpha$ emission line with 
one peak at the line rest wavelength and the other redshifted to about $3.5\times 10^4 \, {\rm 
km\; s^{-1}}$.

\section{Discussion and conclusions} 
\label{sec:dis}

We analysed the three optical spectra of the TDE  candidate PTF09djl, after careful treatment of the TDE 
featureless continuum and host galaxy starlight. The double-peaked H$\alpha$ line profiles are well 
reproduced with an elliptical disk of semimajor about $339.7 r_{\rm S}$ (apocenter $667.9 r_{\rm 
S}$) and eccentricity $0.966$.
The peculiar line substructures with one peak at the line rest wavelength and the other redshifted to 
$\sim 3.5\times 10^4\, {\rm km\; s^{-1}}$ are due to the large disk inclination $88\degr$ and 
pericenter orientation nearly vertical the observer. 

For an accretion disk with  $e_{\rm d} \simeq 0.966$, the conversion efficiency of matter into radiation 
is $\eta \simeq 4.2\times 10^{-3}$. For typical stellar tidal disruptions with $\beta = r_{\rm t}/r_{\rm p}
\simeq 1.48 m_*^{-\zeta+2/3}M_6^{-2/3} 
\simeq 1$, the peak accretion rate is $\dot{M}_{\rm p} \simeq A_{5/3} (M_{\rm BH} /10^6 M_\odot)^{-1/2} 
 M_\odot/ {\rm yr}$ with $A_{5/3} \simeq 1.2$ \citep{gui13}. For a BH of mass $M_{\rm BH} \simeq  
 2.1\times 10^6 M_\odot$, PTF09djl has a peak accretion rate $\dot{M}_{\rm p} \simeq 0.83 M_\odot/{\rm 
 yr}$ and luminosity $L_{\rm p} = \eta \dot{M}_{\rm p} c^2 \simeq 2.0\times 10^{44}\, {\rm erg\; s^{-1}} 
 \simeq 0.75 L_{\rm Edd}$ with $L_{\rm Edd}$ the Eddington luminosity. Accretion disk of sub-Eddington 
 accretion rate is expected to be optically thick and geometrically thin. 

The calculations of radiative transfer show that optical  emission lines are prominent in the 
reflected spectra from an ionized, optically thick accretion disk irradiated by X-rays when the 
ionization parameter is in the range $1 \leq \zeta {\rm (erg \; cm \; s^{-1})} \la 500$ \citep{gar13}, where 
$\zeta = 4\pi F_{\rm x} / n_{\rm e}$ with $F_{\rm x}$ the integrated flux in the energy range $0.1-300\, 
{\rm keV}$ and $n_{\rm e}$ the electron number density. The accretion disk of semimajor $a_{\rm d} 
\simeq 339.7 r_{\rm S}$ and eccentricity $e_{\rm d} \simeq 0.966$ has a typical peak temperature, $T_{\rm p} 
\simeq 4.5\times 10^4 \, {\rm K}$, and is ionized. No hard X-ray observation was made for PTF09djl, but 
the survey of TDE candidates in the {\it Swift} BAT archive show that hard X-ray emission in TDE candidates
should be ubiquitous with luminosity $ L_{\rm x} = (0.3 - 3) \times 10^{44}  \, {\rm ergs \; s^{-1}}$ in the 
energy range $20-195\, {\rm keV}$ \citep{hry16}. Provided a typical X-ray luminosity $L_{\rm x} 
\sim 10^{44}  \, {\rm ergs \; s^{-1}}$ from extended X-ray source corotating with disk below, 
the ionization parameter of PTF09djl is $\zeta \sim 6\pi L_{\rm x} (m_{\rm p}/1.2M_*) (1+e_{\rm d}) 
a_{\rm d} (H/r) [1+ (H/r)^2]^{-1} \simeq 54 \, {\rm erg \; cm \; s^{-1}}$ for $M_* \sim M_\odot$ and disk 
opening angle $H/r\sim 0.1$, where $m_{\rm p}$ is the mass of proton and the typical disk mass 
at peak $M_{\rm d} \sim M_*/3$ is assumed.

Because electron scattering will increase the effective optical depths of emission lines, the ionized disk 
atmosphere becomes effectively optically thick to H and He~II emission lines at different electron scattering 
depths \citep{rot16}. Strong He emission lines are expected because they are produced in a region of 
electron scattering depth a few times larger than that for H lines \citep{rot16}. He emission lines are 
absent/weak in the optical spectra of PTF09djl. It is probably due to the viewing angle effects that the 
effective optical depth changes with disk inclination angle $\tau_{\rm eff} = \tau_{\rm es}/\cos(i_{\rm 
d})$ \citep{gar14}. The accretion disk of PTF09djl is highly inclined with $i_{\rm d} \simeq 88\degr$, and 
the He emission lines originally formed at large $\tau_{\rm es}$ is much more attenuated than H emission 
lines formed near the disk surface. Our results suggest that the 
broad optical emission lines in optical TDEs may originate in the ionized elliptical accretion disk and 
the diversity of line intensity ratios of the species among the optical TDEs is probably due to the 
different disk inclinations.

In conclusion, we have successfully modelled the peculiar double-peaked H$\alpha$ profiles of the TDE 
candidate PTF09djl with a relativistic elliptical disk, without the need to invoke bulk motion, like BH recoil. 
Our results show
that modelling the complex and asymmetric line profiles of TDEs provides a powerful tool to probe the 
structure of their transient accretion disk and potential kinematic signatures of SMBH binaries or recoiling 
SMBHs.


\section*{Acknowledgements}

We are grateful to Iair Arcavi for  providing us the electronic data of the spectra. 
This work is supported by the National Natural Science Foundation of China (NSFC11473003) and the 
Strategic Priority Research Program of the Chinese Academy of Sciences (grant no. XDB23010200 and 
no. XDB23040000). LCH was supported by the National Key R\&D Program of China (2016YFA0400702) 
and the National Science Foundation of China (11473002, 11721303).

\bsp	
\label{lastpage}
\end{document}